\documentclass[US Letter]{article}
\usepackage{spconf,amsmath,graphicx,amssymb,delarray,subfigure,verbatim,booktabs}
\usepackage{diagbox, makecell}
\usepackage{multirow}
\usepackage[english]{babel}
\usepackage{lipsum}
\usepackage{setspace}

\title{Joint magnitude estimation and phase recovery using Cycle-in-Cycle GAN for non-parallel speech enhancement}
\name{Guochen Yu$^{\star \dagger}$\thanks{This work was supported in part by the National Natural Science Foundation of China under Grant 61631016 and in part by the National Key R\&D Program of China under Grant No. SQ2020YFF0426386. Chengshi Zheng is the corresponding author. }, Andong Li$^{\dagger}$, Yutian Wang$^{\star }$, Yinuo Guo$^{\ast}$, Hui Wang$^{\star}$, Chengshi Zheng$^{\dagger}$}
\address{$^{\star}$ State Key Laboratory of Media Convergence and Communication, Communication University \\of China, Beijing, China\\
	$^{\dagger}$ Institute of Acoustics, Chinese Academy of Sciences, Beijing, China\\
	$^{\ast}$  Bytedance, Beijing, China\\
	\{yuguochen, wangyutian, hwang\}@cuc.edu.cn, \{liandong, cszheng\}@mail.ioa.ac.cn }

\begin{document}
	\ninept
	\maketitle
	\begin{abstract}
		For the lack of adequate paired noisy-clean speech corpus in many real scenarios, non-parallel training is a promising task for DNN-based speech enhancement methods. However, because of the severe mismatch between input and target speeches, many previous studies only focus on the magnitude spectrum estimation and remain the phase unaltered, resulting in the degraded speech quality under low signal-to-noise ratio conditions. To tackle this problem, we decouple the difficult target $\emph{w.r.t.}$ original spectrum optimization into spectral magnitude and phase, and a novel Cycle-in-Cycle generative adversarial network (dubbed CinCGAN) is proposed to jointly estimate the spectral magnitude and phase information stage by stage under unpaired data. In the first stage, we pretrain a magnitude CycleGAN to coarsely estimate the spectral magnitude of clean speech. In the second stage, we incorporate the pretrained CycleGAN with a complex-valued CycleGAN as a cycle-in-cycle structure to simultaneously recover phase information and refine the overall spectrum. Experimental results demonstrate that the proposed approach significantly outperforms previous baselines under non-parallel training. The evaluation on training the models with standard paired data also shows that CinCGAN achieves remarkable performance especially in reducing background noise and speech distortion.

		
	\end{abstract}
	
	\vspace{-1mm}
	
	\begin{keywords}
		speech enhancement, non-parallel, Cycle-in-Cycle GAN, magnitude spectrum estimation, phase recovery
	\end{keywords}
	\vspace{-3mm}
	\section{Introduction}
	\vspace{-3mm}
	Speech is often corrupted by various types of background noise in real-world environments. These interferences severely degrade the performance of many speech-related tasks such as automatic speech recognition (ASR) systems or hearing aids~{\cite{loizou2013speech}}.  
	To cope with these degradations, speech enhancement (SE) can be applied to remove or attenuate the background noise for better speech quality and intelligibility. Recent SE approaches based on deep neural networks (DNNs) have demonstrated their superior capability in dealing with non-stationary noise under low signal-to-noise ratio (SNR) conditions~{\cite{wang2018supervised}}. These DNN-based methods can be roughly categorized into masking-based approaches~{\cite{wang2014training}} and mapping-based approaches~{\cite{xu2014regression}}. More recently, generative adversarial networks (GANs)~{\cite{goodfellow2014generative}} begin to gain development in the single-channel SE area~{\cite{pascual2017segan, baby2019sergan, fu2019metricgan, soni2018time, liu2020cp, zhang2020loss, phan2021self}}, in which a generator ($G$) aims to conduct the enhancement process and a discriminator ($D$) aims to distinguish between the generated speech features and real clean ones. For these standard supervised methods, the networks are trained to minimize the discrepancy between the enhanced output and the clean target. 
	
	Despite the impressive performance of the above DNN-based SE approaches, they always need numerous paired clean-noisy samples to conduct supervised training and improve the generalization. However, in some real scenarios, it is troublesome to record parallel clean-noisy pairs, and we can only obtain clean speech that mismatches the source noisy speech. In this case, standard supervised SE methods usually cannot produce satisfactory performance. To tackle this problem, cycle-consistent GAN (CycleGAN) has been developed to conduct unsupervised SE, which was originally used for unpaired image-to-image translation~{\cite{zhu2017unpaired}}. In the SE area, CycleGAN-based approaches demonstrate the remarkable ability in preserving the speech structure and improving speech quality in both paired and unpaired cases~{\cite{meng2018cycle, xiang2020parallel, wang2020improved, yu2021cyclegan,yu2021two}}. Nevertheless, these conventional CycleGAN-based approaches have several limitations for non-parallel SE. First, to ensure cycle consistency of the original noisy speech and target clean speech for unpaired data, the enhanced signal always contains the original noise information. In other words, cycle-consistency-based methods remain audible residual noise, which is challenging to be eliminated completely. Second, previous CycleGAN-based SE methods only focus on magnitude spectrum estimation and remain the noisy phase unaltered. This is because simultaneously restoring the clean spectral magnitude and phase information still remains to be a tough task, let alone when noisy-clean pairs are mismatched. However, recent studies revealed the importance of the accurate phase for improving perceptual speech quality, especially in low SNR cases~{\cite{paliwal2011importance}}.
	
	\begin{figure*}[ht]
		\centering
		\centerline{\includegraphics[width=2\columnwidth]{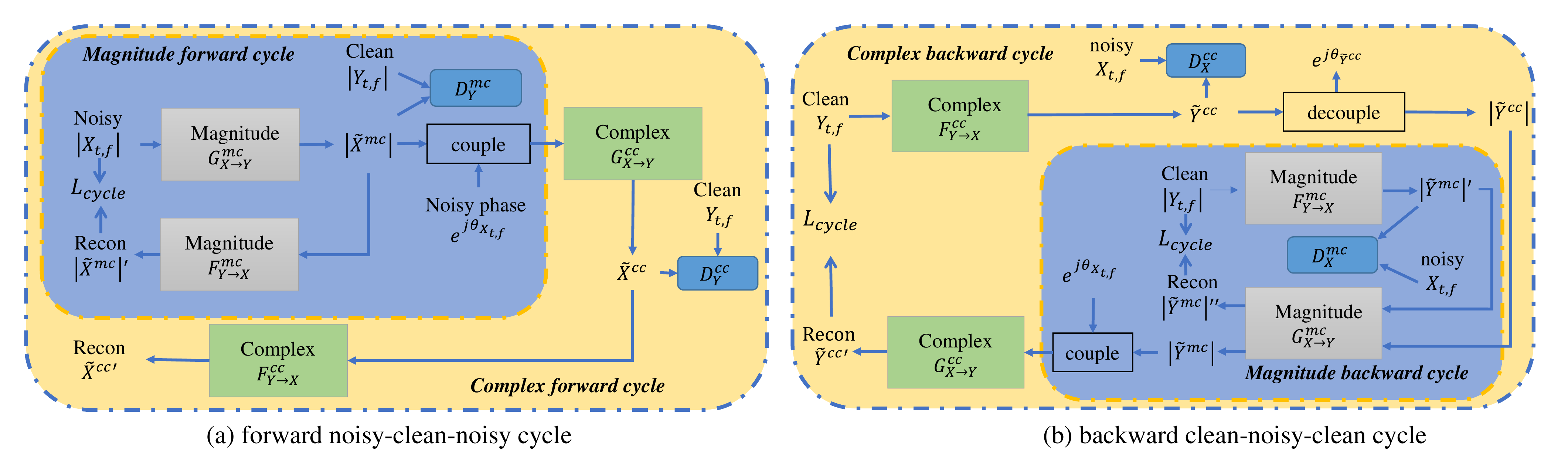}}
		\vspace{-6mm} 
		\caption{
			System flowchart of CinCGAN. The magnitude and complex cycle are shown in the yellow and blue dotted boxes, respectively. (a) The diagram of the forward cycle in CinCGAN. (b) The diagram of the backward cycle in CinCGAN.
		}
		\label{fig:architecture}
		\vspace{-6mm}
	\end{figure*}
	Multi-stage algorithms based on curriculum learning begin to thrive in the SE area~{\cite{li2021two,li2021simultaneous}}, where the original difficult task is decomposed into multiple simpler sub-tasks and a better result can be induced progressively. Motivated by these studies, we couple a magnitude spectrum estimation CycleGAN (MCGAN) and a complex spectrum refined CycleGAN (CCGAN) as a Cycle-in-Cycle GAN (CinCGAN) paradigm to estimate the clean spectral magnitude and phase information step-by-step under non-parallel training. In the first step, the complex original spectrum is decoupled into the spectral magnitude and phase, and only the magnitude of the clean spectrum is estimated by MCGAN. Subsequently, a complex-valued CCGAN is integrated with the pretrained MCGAN to jointly recover the clean spectral magnitude and phase information. To be specific, we couple the estimated magnitude spectrum by MCGAN and original phase as a coarse-estimated complex spectrum, and feed it into CCGAN to estimate both real and imaginary (RI) components of the clean spectrum. 
	
	The remainder of the paper is organized as follows. In Section~{\ref{Sec2}}, the overall diagram is described in detail. The experimental setup is presented in Section~{\ref{Sec3}}, while the results and analysis are given in Section~{\ref{Sec4}}. Some conclusions are drawn in Section~{\ref{Sec5}}.
	\vspace{-2mm}
	\section{Algorithm description\label{Section2}}
	\label{Sec2}
	\vspace{-2mm}  	
	\subsection{ Cycle-in-Cycle Scheme \label{Section21}}
	\vspace{-1mm} 
	To simultaneously estimate magnitude and phase information of clean spectrum under non-parallel training conditions, we integrate a magnitude spectrum estimation CycleGAN (MCGAN) with a complex spectrum restoration CycleGAN (CCGAN) as a Cycle-in-Cycle GAN (CinCGAN). The proposed CinCGAN consists of a forward noisy-clean-noisy cycle and a backward clean-noisy-clean cycle, as shown in Fig.~{\ref{fig:architecture}}. In the forward cycle, the enhancement procedure can be divided into two steps. First, we decouple the complex spectrum into spectral magnitude and phase, and only the amplitude is processed. A magnitude generator $G^{mc}_{X\rightarrow Y}$ is utilized to convert the compressed noisy spectral magnitude $\left | {X}_{t,f} \right |$ to the "clean" one $|\widetilde{X}^{mc} |$, while a magnitude discriminator $D^{mc}_{Y}$ is used to distinguish the generated spectral magnitude (\emph{i.e.}, $| \widetilde{X}_{t,f} |$) from the clean one (\emph{i.e.}, $\left | {Y}_{t,f} \right |$), where $t$ and $f$ denote the time and frequency indices. To constrain the cycle consistency under unpaired data, an inverse generator $F^{mc}_{Y\rightarrow X}$ is employed to reconstruct the noisy spectral magnitude from the generated one. Subsequently, we couple the estimated spectral magnitude and the original phase (\emph{i.e.}, $e^{j\theta_{X_{t,f}}}$) into a coarsely estimated complex spectrum and feed it into CCGAN. By estimating both RI parts of the clean spectrum, CCGAN can refine the spectral magnitude and recover the clean phase. Specifically, a complex-valued generator $G^{cc}_{X\rightarrow Y}$ attempts to generate the enhanced complex spectrum $\widetilde{X}^{cc}$ as clean one, and a complex discriminator $D^{cc}_{Y}$ attempts to discriminate between the generated and clean complex spectrum (\emph{i.e.}, $ {Y}_{t,f}$). Similarly, an inverse complex generator $F^{cc}_{Y\rightarrow X}$ brings the enhanced complex spectrum $\widetilde{X}^{cc}$ back to the original noisy domain. To better constrain the mapping optimization space, a backward clean-noisy-clean cycle is also designed and jointly trained with the forward cycle. In the testing stage, the whole enhancement procedure can be formulated as:
	\vspace{-1mm}
	\begin{gather}
		\label{eqn3}
		|\widetilde{X}^{mc}| = G^{mc}_{X\rightarrow Y}\left( |X_{t,f}|; \Phi_{1} \right),\\
		\widetilde{X}^{mc} = |\widetilde{X}^{mc}| \exp(j\theta_{{X}_{t,f}}), \\
		\widetilde{X}^{cc} = G^{cc}_{X\rightarrow Y} \left( \widetilde{X}^{mc}; \Phi_{2} \right),
		\vspace{-3mm}
	\end{gather}
	where $|X_{t,f}|$ and $\theta_{{X}_{t,f}}$ denote the noisy spectral magnitude and phase, respectively. $|\widetilde{X}^{mc}|$ and $\widetilde{X}^{cc}$ denote the outputs of the noisy-to-clean magnitude generator $G^{mc}_{X\rightarrow Y}$ and complex generator $G^{cc}_{X\rightarrow Y}$ with parameter set $\Phi_{\left(\cdot\right)}$.

	\vspace{-3mm} 
	\subsection{Network Architecture\label{Section24}}
\begin{figure}[t]
		\centering
		\centerline{\includegraphics[width=1\columnwidth]{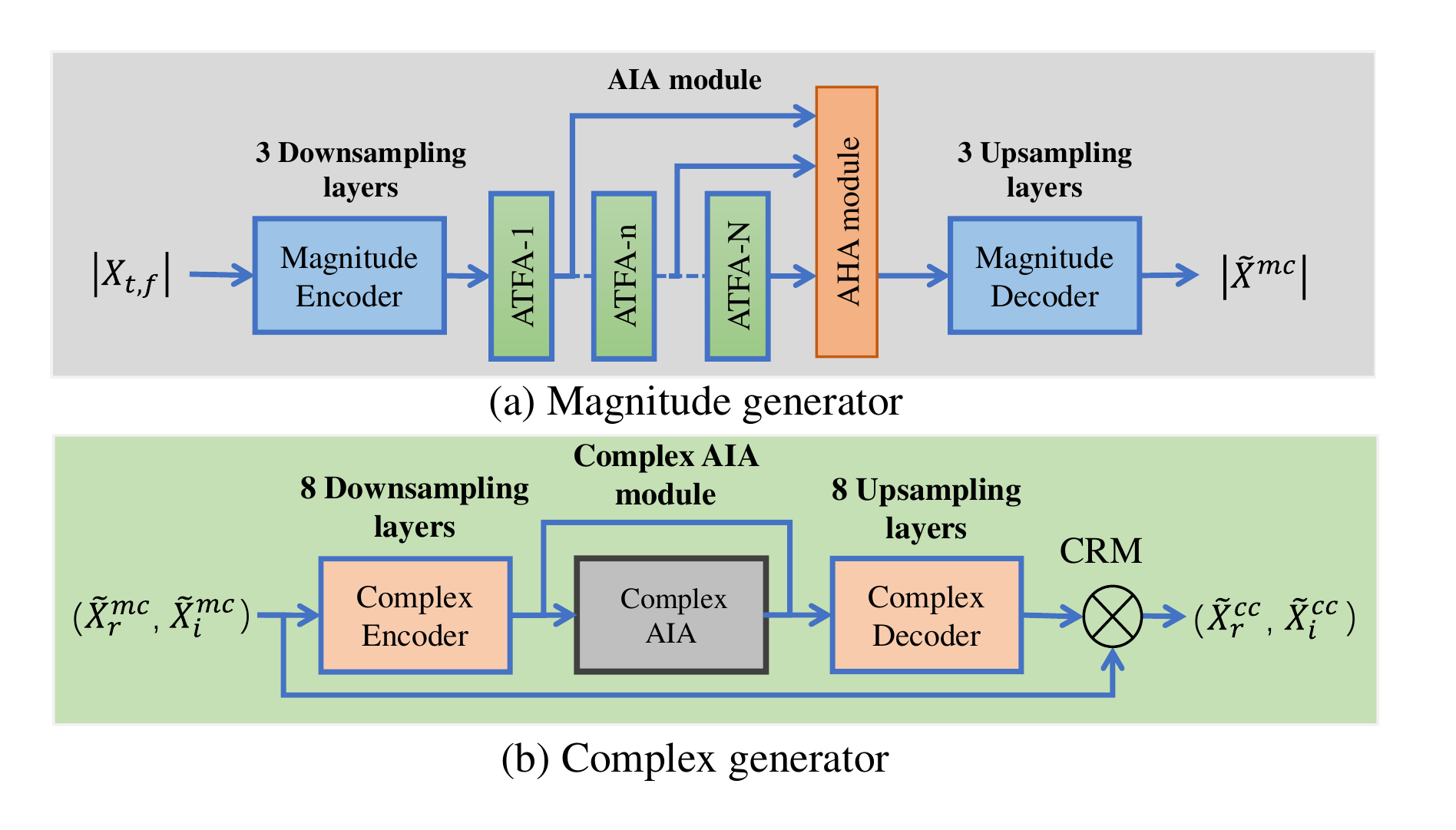}}
		\vspace{-6mm}
		\caption{
			(a) The diagram of the magnitude generators in MCGAN. (b) The diagram of the complex generators in CCGAN.
		}
		\label{fig:generators}
		\vspace{-6mm}
	\end{figure}

	\vspace{-1mm}  
	Following the same configuration as~{\cite{yu2021cyclegan}}, the magnitude generator in MCGAN consists of three major parts, including three downsampling layers as the encoder, an adaptive attention-in-attention (AIA) module and three upsampling layers as the decoder, as shown in Fig.~{\ref{fig:generators}} (a). Each downsampling/upsampling layer block is composed of a 2D (de)convolution layer, followed by instance normalization (IN), Parametric Relu activation function (PReLU) and gated liner units (GLUs). The proposed AIA module consists of six adaptive time-frequency attention (ATFA) modules and an adaptive hierarchical attention (AHA) module, as proposed in~{\cite{yu2021cyclegan}}. More specifically, ATFA consists of two branches: an adaptive time attention branch (ATAB) and an adaptive frequency attention branch (AFAB). The two branches cooperate to capture the global dependencies along temporal and spectral axes in parallel. The AHA module integrates different hierarchical feature maps given all outputs of ATFA modules and fuses the global context of all intermediate outputs with different weights, thus helping to guide the feature learning progressively. The output channels of each layer in the encoder are (16, 32, 64), while the kernel size and strides are set to (3,~5) and (1,~2) along the time and frequency axes, respectively.  
	
	We also design the pure complex-valued generators of CCGAN similar as~{\cite{yu2021two}}, which consists of eight complex-valued encoder/decoder blocks, and an complex-valued AIA module, as shown in Fig.~{\ref{fig:generators}}(b). The complex encoder/decoder blocks include complex-valued 2D (de)convolutions, complex IN and real-valued PReLU. The complex-valued convolution simulates the correlation between real and imaginary parts of feature maps by complex multiplication, which is composed of four traditional convolution operations~{\cite{trabelsi2017deep}}. To further capture long-term dependency with complex-valued features, we utilize the above attention-in-attention mechanism in the complex-valued manner. The number of the channels for each complex encoder layer is (32, 32, 64, 64, 128, 128, 256, 256), with kernel size and strides set to (3,~3) and (1,~2), respectively. 
	
	We utilize multi-scale discriminators that can force the generator to produce feature maps with global consistency and finer details, as pointed out in ~{\cite{yu2021cyclegan}}. Both the magnitude and complex discriminators are composed of six real-valued 2D convolutions, spectral normalization (SN) and PReLU. The kernel size for each 2D convolutions layer is (3,~5) except (1,~1) for the last layer in the temporal and frequency axis, respectively. The number of channels throughout the 2D convolutions is (32, 32, 64, 64, 128, 1), with stride set to (1,~2). Note that we concatenate RI components along the channel axis and feed them into the complex discriminator.
	
	\vspace{-4mm}
	\subsection{Loss Function \label{Section23}}
	\vspace{-2mm} 
	In both MCGAN and CCGAN, we use the following three losses to jointly optimize the forward and backward cycle, namely relativistic adversarial loss, cycle-consistency loss, and an identity mapping loss. Taking the magnitude CycleGAN (MCGAN) as an example, for the forward noisy-to-clean mapping, the relativistic average least-square (RaLS) adversarial loss~{\cite{jolicoeur2018relativistic}} is used to enforce the enhanced magnitude spectra indistinguishable from the clean ones, which is calculated as follows. 
	\vspace{-2mm}
	\begin{gather}	
		\mathcal{L}_{Radv}(D^{mc}_Y)= \mathbb{E}_{y}\left [(D^{mc}_Y(y)-\mathbb{E}_{x}D^{mc}_Y(G_{X\rightarrow Y}(x))-1)^{2}  \right ] +  \notag \\ \mathbb{E}_{x}\left [(D^{mc}_Y(G_{X\rightarrow Y}(x))-\mathbb{E}_{y}(D^{mc}_Y(y))+1)^{2}  \right ]\notag, \\	
		\mathcal{L}_{Radv}(G^{mc}_{X\rightarrow Y})=  \mathbb{E}_{x}\left [(D^{mc}_Y(G_{X\rightarrow Y}(x)))-\mathbb{E}_{y}D^{mc}_Y(y) -1)^{2} \right] \notag \\+ \mathbb{E}_{y}\left [(D^{mc}_Y(y)-\mathbb{E}_{x}(D^{mc}_Y(G_{X\rightarrow Y}(x)))+1)^{2}  \right ],
		\vspace{-2mm} 
	\end{gather}
	
	Analogously, $\mathcal{L}_{Radv}(D^{mc}_X)$ and $\mathcal{L}_{Radv}(F^{mc}_{ Y\rightarrow X})$ are employed in the inverse magnitude clean-to-noisy mapping. The cycle-consistency loss and identity mapping loss are utilized to constrain the mapping functions under non-parallel training conditions and to further preserve the linguistic composition of the original speech, which can be defined as:
	\vspace{-2mm}
	\begin{gather}	
		\mathcal{L}_{cycle}(G^{mc}_{X\rightarrow Y},F^{mc}_{ Y\rightarrow X})=\mathbb{E}_{x}\left [\left \|F_{ Y\rightarrow X}(G^{mc}_{X\rightarrow Y}(x)) -x \right \|_{1}\right ] \notag \\+\mathbb{E}_{y}\left [\left \|G^{mc}_{X\rightarrow Y}(F^{mc}_{ Y\rightarrow X}(y)) -y \right \|_{1}\right ]\tag{5},\\
		\mathcal{L}_{id}(G^{mc}_{X\rightarrow Y},F^{mc}_{ Y\rightarrow X})=\mathbb{E}_{x}\left [\left \|F^{mc}_{ Y\rightarrow X}(x) -x \right \|_{1}\right ]+ \notag \\ \mathbb{E}_{y}\left [\left \|G^{mc}_{X\rightarrow Y}(y) -y \right \|_{1}\right ],\tag{6}
		\vspace{-2mm} 
	\end{gather}
	\vspace{-4mm} 
	
	In the first step, we pretrain MCGAN alone with $\mathcal{L}_{MCGAN}$ until convergence, which is calculated as:
	\vspace{-2mm}
	\begin{equation}
		\begin{gathered}	
			\mathcal{L}_{MCGAN}=\mathcal{L}_{Radv}(G^{mc}_{X\rightarrow Y},D^{mc}_{Y}) + \mathcal{L}_{Radv}(F^{mc}_{Y\rightarrow X},D^{mc}_{X}) \\ +\lambda _{cycle}\mathcal{L}_{cycle}(G^{mc}_{X\rightarrow Y},F^{mc}_{Y\rightarrow X}) +\lambda _{id}\mathcal{L}_{id}(G^{mc}_{X\rightarrow Y},F^{mc}_{Y\rightarrow X}) \tag{7}, \\
		\end{gathered}
	\end{equation}
	where $\lambda_{cycle}$ and $\lambda_{id}$ are empirically set to be 5 and 10, respectively. Then, we jointly fine-tune MCGAN and CCGAN with the same relativistic adversarial loss, cycle-consistency loss, and identity mapping loss, where the parameters of the first sub-network are initialized with the pretrained MCGAN. The overall loss functions of CinCGAN are expressed as:
	\vspace{-2mm}
	\begin{equation}
		\begin{gathered}
			\mathcal{L}_{CinCGAN}= \gamma\mathcal{L}_{MCGAN} +\mathcal{L}_{CCGAN}\tag{8} \\
		\end{gathered}
	\end{equation}
	where $\mathcal{L}_{CCGAN}$ denotes the loss function for the complex cycle. With some trials, we empirically set $\gamma= 0.1$ in all the following experiments .

	\section{Experiments\label{Section3}}
	\label{Sec3}
	\vspace{-3mm} 
	\subsection{Datasets\label{Section31}}
	\vspace{-2mm} 
	The dataset used in this work is publicly available as proposed in~{\cite{valentini2016investigating}}, which is a selection of the Voice Bank corpus~{\cite{veaux2013voice}} with 28 speakers for training and another 2 unseen speakers for testing. The training set consists of 11,572 mono audio samples, while the test set contains 824 utterances by 2 speakers (one male and one female). For the training set, audio samples are mixed together with one of the 10 noise types (\emph{i.e.}, two artificial (babble and speech shaped) and eight real noise from the DEMAND database~{\cite{thiemann2013diverse}}) at four SNRs of $\left\{ 0\rm{dB}, 5\rm{dB}, 10\rm{dB}, 15\rm{dB}\right\}$. The testing utterances are created with 5 unseen test-noise types (all from the DEMAND database) at SNRs of $\left\{2.5\rm{dB}, 7.5\rm{dB}, 12.5\rm{dB}, 17.5\rm{dB} \right\}$. The original raw waveforms are downsampled from 48kHz to 16kHz for our task.

	\vspace{-4mm}
	\subsection{Implementation Setup\label{Section32}}
	\vspace{-2mm}  
	The Hanning window of length 32ms is applied, with 75\% overlap between adjacent frames. The 512-point STFT is utilized, leading to a 257-D spectral feature. Due to the efficacy of power compression in dereverberation and denoising tasks~{\cite{li2021simultaneous}}, we conduct the power compression toward the spectral magnitude while leaving the phase unaltered, and the optimal compression coefficient is set to 0.5. For the non-parallel training strategy, we randomly crop a fixed-length segment (\emph{i.e.}, 108 frames) from a randomly selected noisy audio file as the input, while the target is a randomly selected clean audio file whose content is totally different from the input audio. We adopt the Adam optimizer~{\cite{kingma2014adam}} with the momentum term $\beta_{1}=0.9$, $\beta_{2}=0.999$. We divide our training process into two steps. Firstly, we pretrain the MCGAN with an initial learning rate of 0.0002 for discriminators and 0.0005 for generators, respectively. Subsequently, we jointly fine-tune the pretrained MCGAN and CCGAN as the proposed CinCGAN with a learning rate of 0.0002 for all generators and discriminators, while the learning rate is linearly decayed after 40 epochs. We use $\mathcal{L}_{id}$ in each cycle only for the first 20 epochs. The sound demo samples are available online~{\footnote{https://github.com/yuguochencuc/CinCGAN-SE}}.
	\vspace{-3mm}

	\section{Results and Analysis\label{Section4}}
	\label{Sec4}
	\vspace{-2mm}  
	
	\renewcommand\arraystretch{0.88}
	\begin{table*}[t!]
		\caption{Experimental results among different models including GAN-based methods and the proposed CycleGAN-based methods under unpaired data. Best scores are highlighted in \textbf{bold}. }
		\tiny
		\label{results}
		\centering 
		\small
		\scalebox{0.88}{
			\begin{tabular}{l|l|cc|cc|cccccccc}
				\hline
				\multirow{2}*{\textbf{Methods}}&\multirow{2}*{\textbf{Feature type}} & \multicolumn{2}{c|}{\textbf{Magnitude}}& \multicolumn{2}{c|}{\textbf{Complex }} &\multirow{2}*{\textbf{PESQ}}  &\multirow{2}*{\textbf{STOI(\%)}}   &\multirow{2}*{\textbf{CSIG}}  &\multirow{2}*{\textbf{CBAK}}  &\multirow{2}*{\textbf{COVL}}  &\multirow{2}*{\textbf{SegSNR}}  &\multirow{2}*{\textbf{DNSMOS}} \\ 
				\cline{3-6}   & & fc & bc & fc & bc \\ \hline
				\multicolumn{1}{l|}{Unprocessed} &{\makecell[c]{--}} &{\makecell[c]{--}} &{\makecell[c]{--}} &{\makecell[c]{--}} &{\makecell[c]{--}} & 1.97 & 92.1 & 3.35 & 2.44 & 2.63 &1.68 & 3.02\\ \hline
				\multicolumn{12}{c}{\textbf{GAN-based methods}} \\ \hline
				\multicolumn{1}{l|}{MGAN} & Magnitude& $\times$ & $\times$ &$\times$  & $\times$&2.03 & 91.6 & 3.54 & 2.78 & 2.72 & 5.28 & 2.72 \\ 
				\multicolumn{1}{l|}{MGAN+fc}& Magnitude & $\checkmark$ &  $\times$& $\times$ &$\times$ &2.58 & 92.8 & 3.81 & 3.03 & 3.19 & 5.28 & 3.26 \\ 
				\multicolumn{1}{l|}{CGAN }& RI components& $\times$ &$\times$ &$\times$ & $\times$& 1.86 & 88.9 & 3.17 & 2.62 & 2.64 & 2.98 & 2.63 \\ 
				\multicolumn{1}{l|}{CGAN+fc}& RI components&$\times$ & $\times$ &$\checkmark$ & $\times$& 2.32 & 91.2 & 3.48 & 2.74 & 3.18 & 4.67 & 3.04 \\ \hline
				\multicolumn{12}{c}{\textbf{Proposed CycleGAN-based Systems}} \\ \hline
				\multicolumn{1}{l|}{MCGAN}& Magnitude & $\checkmark$&$\checkmark$ & $\times$&$\times$ & 2.67 & 93.2 & 3.86 & 3.20 & 3.21 &7.23 &3.47 \\ 
				\multicolumn{1}{l|}{CCGAN}& RI components &$\times$ & $\times$&$\checkmark$& $\checkmark$& 2.56 & 92.1 & 3.67 & 3.10 & 3.16 & 5.38 &3.42 \\ 			
				\multicolumn{1}{l|}{CinCGAN (I) }& Magnitude + RI  & $\checkmark$& $\times$ &$\checkmark$ &$\times$ & 2.70 & 93.4 & 3.93 & 3.24 & 3.25  &7.34 &3.44 \\			
				\multicolumn{1}{l|}{CinCGAN (II) }& Magnitude + RI  & $\checkmark$& $\checkmark$& $\checkmark$& $\times$& 2.77 & 93.6 & 3.96 & 3.02 & 3.30  &4.49 & 3.49\\			
				\multicolumn{1}{l|}{CinCGAN (III) } & Magnitude + RI  & $\checkmark$& $\times$& $\checkmark$ & $\checkmark$& 2.73 & 93.5 & 3.94 & 3.27 & 3.29 & 7.98 & 3.51 \\
				\multicolumn{1}{l|}{CinCGAN (IV) }& Magnitude + RI & $\checkmark$& $\checkmark$& $\checkmark$& $\checkmark$& \textbf{2.84} & \textbf{94.1} & \textbf{4.10} & \textbf{3.36} & \textbf{3.37} &\textbf{8.91} &\textbf{3.53}\\\hline			
				
			\end{tabular}
		}
		\vspace{-6mm}
	\end{table*}

	\renewcommand\arraystretch{0.91}
	\begin{table}[t!]
		\vspace{-1mm}
		\caption{Comparison with other GAN and Non-GAN based systems under standard paired data. "$-$" denotes that the result is not provided in the original paper.}
		\label{tbl:VB-results}
		\centering
		\small
		\scalebox{0.70}{
			\begin{tabular}{lcccccc}
				\hline
				\multicolumn{1}{l|}{\textbf{Methods}} &\multicolumn{1}{l|}{\textbf{Feature type}} & \textbf{PESQ} & \textbf{STOI(\%)}  & \textbf{CSIG} & \textbf{CBAK} & \textbf{COVL} \\ \hline
				\multicolumn{1}{l|}{Unprocessed} & \multicolumn{1}{l|}{\makecell[c]{--}} & 1.97 & 92.1 & 3.35 & 2.44 & 2.63 \\ \hline
				\multicolumn{6}{c}{\textbf{GAN-based Systems}} \\ \hline
				\multicolumn{1}{l|}{SEGAN~{\cite{pascual2017segan}} } & \multicolumn{1}{l|}{Waveform} & 2.16 & 92.5 & 3.48 & 2.94 & 2.80 \\ 
				\multicolumn{1}{l|}{MMSEGAN~{\cite{soni2018time}} } &\multicolumn{1}{l|}{Gammatone } & 2.53 & 93.0 & 3.80 & 3.12 & 3.14 \\ 
				\multicolumn{1}{l|}{SERGAN~{\cite{baby2019sergan}} }  & \multicolumn{1}{l|}{Waveform} & 2.51 & 93.7 & 3.78 & 3.23 & 3.16 \\ 
				\multicolumn{1}{l|}{CP-GAN~{\cite{liu2020cp}} } &  \multicolumn{1}{l|}{Waveform} & 2.64 & 94.0 & 3.93 & 3.29 & 3.28 \\ 			
				\multicolumn{1}{l|}{MetricGAN~{\cite{fu2019metricgan}} } &\multicolumn{1}{l|}{Magnitude} & 2.86 & \makecell[c]{--} & 3.99 & 3.18 & 3.42 \\ 
				\multicolumn{1}{l|}{CRGAN~{\cite{zhang2020loss}}} &\multicolumn{1}{l|}{Magnitude} & \textbf{2.92} & 94.0 & 4.16 & 3.24 & \textbf{3.54} \\ 
				\multicolumn{1}{l|}{SASEGAN~{\cite{phan2021self}} } &\multicolumn{1}{l|}{Waveform} & 2.36 & 93.5 & 3.54 & 3.08 & 2.93\\ \hline
				\multicolumn{6}{c}{\textbf{Non-GAN based Systems}} \\ \hline			\multicolumn{1}{l|}{Wave-U-net~{\cite{stoller2018wave}} } & \multicolumn{1}{l|}{Waveform} & 2.64 & \makecell[c]{--} & 3.56 & 3.08 & 3.09 \\  
				\multicolumn{1}{l|}{DFL-SE~{\cite{germain2019speech}} } &  \multicolumn{1}{l|}{Waveform} &\makecell[c]{--} & \makecell[c]{--} & 3.86 & 3.33 & 3.22 \\ 
				\multicolumn{1}{l|}{CRN-MSE~{\cite{tan2018convolutional}} } & \multicolumn{1}{l|}{Magnitude} & 2.61 & 93.8 & 3.78 & 3.11 & 3.24 \\ 
				\multicolumn{1}{l|}{GCRN~{\cite{tan2019learning}} } & \multicolumn{1}{l|}{RI components} & 2.51  & 94.0 & 3.71 & 3.24 & 3.09 \\ 			
				\multicolumn{1}{l|}{DCCRN~{\cite{hu2020dccrn}} } & \multicolumn{1}{l|}{RI components} & 2.68 & 93.9 & 3.88 & 3.18 & 3.27 \\ 
				
				\multicolumn{1}{l|}{TFSNN~{\cite{yuan2020time}} } & \multicolumn{1}{l|}{Waveform} & 2.79 & \makecell[c]{--} & 4.17 & 3.27 & 3.49 \\ \hline
				
				\multicolumn{6}{c}{\textbf{Proposed CycleGAN-based approaches}} \\ \hline
				\multicolumn{1}{l|}{MCGAN} &  \multicolumn{1}{l|}{Magnitude}  & 2.74 & 93.6 & 3.96 & 3.25 & 3.29 \\ 
				\multicolumn{1}{l|}{CCGAN} &  \multicolumn{1}{l|}{RI components} & 2.60 &92.8 & 3.82 & 3.12 & 3.20 \\ 
				\multicolumn{1}{l|}{CinCGAN} &  \multicolumn{1}{l|}{Magnitude+ RI} & 2.86 & \textbf{94.4} & \textbf{4.18} & \textbf{3.38} & 3.42 \\ \hline
			\end{tabular}
		}
		\vspace{-6mm}
	\end{table}
	
	We use the following objective metrics to evaluate speech enhancement performance: the perceptual evaluation of speech quality (PESQ)~{\cite{rix2001perceptual}}, short-time objective intelligibility (STOI)~{\cite{taal2010short}}, segmental signal-to-noise ratio (SegSNR), and three mean opinion score (MOS) prediction (\emph{i.e.}, signal distortion evaluation (CISG), the intrusiveness of background noise (CBAK) and overall effect (COVL))~{\cite{hu2007evaluation}}. We also evaluate the subjective quality by DNSMOS~{\cite{reddy2020dnsmos}}, which is a robust non-intrusive perceptual speech quality metric. Higher values of all metrics indicate better performance.
	
	\vspace{-4mm}
	\subsection{Comparison with GAN-based and CycleGAN-based methods under non-parallel Training\label{Section42}}
	\vspace{-1mm}  
	To investigate the effectiveness of our proposed method under a non-parallel training strategy, we re-implement two GAN based SE approaches with spectral magnitude and complex spectrum as input (\emph{i.e.}, MGAN and CGAN), and their corresponding models with a forward cycle (fc) (\emph{i.e.}, MGAN+fc and CGAN+fc). We then investigate the performance of different cycle structures, including MCGAN, CCGAN fed with the original noisy complex spectrum, and CinCGANs with different cycles (\emph{i.e.}, CinCGAN (I)-(IV)). To be specific, CinCGAN (I) employs the magnitude and complex fc, while CinCGAN (II) and CinCGAN (III) add the backward cycle in MCGAN and CCGAN, respectively. Furthermore, CinCGAN (IV) employs all the forward and backward cycle in the magnitude and complex domain.
	From Table~{\ref{results}}, one can have the following observations. First, MGAN+fc outperforms other GAN-based methods in terms of all metrics. This fully validates the efficacy of fc under a non-parallel training strategy. Second, when adding a backward cycle (bc) as standard CycleGANs, MCGAN and CCGAN consistently surpass MGAN+fc and CGAN+fc, respectively. Even only estimating the spectral magnitude and remaining the noisy phase unaltered, MCGAN achieves better performance than CCGAN, which aims to estimate both RI components of the clean spectrum. The possible reason is that it seems quite difficult to simultaneously restore the magnitude spectrum and phase information under unpaired data by a one-stage strategy, due to the severe mismatch between input and target. Third, by integrating CCGAN with pretrained MCGAN, CinCGANs with different cycles improve speech performance at different degrees. When only using spectral magnitude fc and complex fc as shown in Figure.1 (a), CinCGAN (I) outperforms MCGAN and CCGAN in terms of most metrics. This phenomenon reveals the significance of simultaneously recovering spectral magnitude and phase information under non-parallel training conditions. That is to say, when the phase is incorporated in the complex cycle, better speech quality can be derived. When adding the spectral magnitude bc, CinCGAN (II) further achieves average 0.07 PESQ, 0.2\% STOI, and 0.05 DNSMOS improvements than CinCGAN (I), which demonstrates the merit of bc in improving perceptual speech quality and intelligibility. However, the spectral magnitude bc forces the noisy-to-clean generators in MCGAN to preserve original speech information, thus leading to residual noise. We also notice that using both spectral magnitude bc and complex bc can facilitate all generators, achieving notable improvements. For example, going from CinCGAN (I) to CinCGAN (IV), average 0.14, 0.7\%, 1.57dB and 0.11 improvements are observed in terms of PESQ, STOI, SegSNR and DNSMOS, respectively. This indicates that CinCGAN with both cycles in MCGAN and CCGAN consistently improves speech quality, while effectively reducing residual noise and speech distortion.

	\vspace{-6mm}
	\subsection{Comparison with other competitive methods under standard parallel training\label{Section43}}
	\vspace{-3mm}
	As presented in Table 2, our proposed methods are compared with other state-of-the-art (SOTA) SE approaches under standard paired data, including GAN-based methods (\emph{e.g.}, SEGAN~{\cite{pascual2017segan}}, MMSEGAN~{\cite{soni2018time}}, SERGAN~{\cite{baby2019sergan}}, MetricGAN~{\cite{fu2019metricgan}}, SASEGAN~{\cite{phan2021self}} and CRGAN~{\cite{zhang2020loss}}) and Non-GAN based methods (\emph{i.e.}, Wave-U-net~{\cite{stoller2018wave}}, DFL-SE~{\cite{germain2019speech}}, CRN-MSE~{\cite{tan2018convolutional}}, GCRN~{\cite{tan2019learning}}, DCCRN~{\cite{hu2020dccrn}},  and TFSNN~{\cite{yuan2020time}}). Note that we re-implement CRN-MSE, GCRN and DCCRN on Voice Bank + DEMAND dataset, and directly use the reported scores of other methods in their original papers. One can have the following observations. Compared with the existing competitive GAN-based methods, CinCGAN provides better performance on STOI, CSIG and CBAK scores, which verifies the merit of CinCGAN in reducing speech distortion and suppressing residual noise components. Although CRGAN achieves a better PESQ score than CinCGAN, which is because that PESQ is directly used as the metric loss for CRGAN, CinCGAN provides average 0.4\%, 0.02 and 0.14 improvements than CRGAN with respect to STOI, CSIG and CBAK, respectively. When compared with Non-GAN based methods, CinCGAN also provides competitive performance in terms of all metrics. For example, CinCGAN achieves average 0.18, 0.7\% ,0.30 0.20 and 0.15 improvements than DCCRN in PESQ, STOI, CSIG, CBAK and COVL, respectively. 
	\vspace{-3mm}

	\vspace{-2mm}
	\section{Conclusions\label{Section5}}
	\label{Sec5}
	\vspace{-3mm}
	In this paper, we propose a novel Cycle-in-Cycle GAN dubbed CinCGAN to jointly recover the spectral magnitude and phase information of clean speech for non-parallel speech enhancement. In the first stage, a standard magnitude spectrum estimation CycleGAN is pretained to only estimate the spectral magnitude under unpaired data. In the second stage, we couple the pretrained CycleGAN with a complex-valued CycleGAN as a cycle-in-cycle structure to further refine the magnitude spectrum and recover phase information. More specifically, the CinCGAN aims at estimating both real and imaginary components of the clean speech complex spectrum. Comprehensive experiments on Voice Bank + DEMAND dataset demonstrate that the proposed approach consistently outperforms original GAN-based and CycleGAN-based baselines under mismatched noisy-clean pairs. When experiments are conducted on standard parallel data, the proposed approach also shows its effectiveness in improving speech quality and reducing speech distortion.
	
	\vspace{-3mm}

	\bibliographystyle{IEEEbib}
	\begin{spacing}{0.85} 
		\bibliography{myrefs} 
	\end{spacing} 
	
\end{document}